\documentclass[aps,seceqnum,preprintnumbers]{revtex4}
\usepackage[dvipdfmx]{graphicx}
\usepackage{amsmath, amssymb}
\usepackage{comment}
\usepackage{float}

\makeatletter
    
    \@addtoreset{equation}{section}
\makeatother

\newcommand{\dd}{\ell}

\begin{document}

\preprint{YITP-14-5}

\title{Mapping de Rham-Gabadadze-Tolley 
bigravity into braneworld setup}

\author{Yasuho Yamashita}
\author{Takahiro Tanaka}

\affiliation{Yukawa Institute for Theoretical Physics,
Kyoto University, 606-8502, Kyoto, Japan}

\begin{abstract}
We discuss whether or not 
bigravity theory can be embedded into the braneworld setup. 
As a candidate, we consider Dvali-Gabadadze-Porrati 
two-brane model with the 
Goldberger-Wise radion stabilization. We will show that 
we can construct a ghost free model whose low energy spectrum 
is composed of 
a massless graviton and a massive graviton with a small mass. 
As is expected, the behavior of this effective theory is 
shown to be identical to de Rham-Gabadadze-Tolley 
bigravity. Unfortunately, this correspondence breaks down 
at a relatively low energy due to the limitation of 
the adopted stabilization mechanism. 
\end{abstract}

\maketitle

\section{Introduction}

Recently, a framework of covariant bimetric gravity with no
ghost has been established~\cite{HR1}.
Gravitational theory with two interacting metrics 
in general suffers from an unwanted ghost degree of freedom, 
called Boulware-Deser(BD) ghost~\cite{BD}.
The finding was that the ghost-free condition is satisfied 
by assuming a restricted form of interaction between two metrics, 
which is called de Rham-Gabadadze-Tolley (dRGT) bigravity~\cite{dRGT1, dRGT2, HR2}.
This interaction has only few parameters 
and makes one of the two graviton modes massive.
In dRGT bigravity, properties of the cosmological 
and/or black-hole solutions have 
been already studied~\cite{cos1, cos2, cos3, cos4, cos5, BH1, BH2, GW}. 
However, it is difficult to have physical intuition about 
these properties. This is partly because 
the form of interaction between two metrics was technically derived so 
as to erase the ghost. 
Furthermore, it is not so clear whether or not 
this theory can be derived from a more natural setup as a 
low energy effective theory. 

In order to improve our understanding of dRGT bigravity, 
we consider to reproduce this model in a braneworld setup as a 
low energy effective theory. 
Two metrics in dRGT bigravity will be identified with 
the metrics induced on two branes embedded 
in higher dimensional bulk spacetime. 
The toy models with two branes placed in five dimensional bulk
have been rather extensively studied~\cite{RS1, RS2}.  
In order to realize a model that effectively includes only 
two gravitons, the mass hierarchy between the lowest 
massive Kaluza-Klein (KK) graviton and the other massive ones will be required.
The eigenvalue problem that determines the 4-dimensional mass spectrum 
in the model with a compact extra dimension is
analogous to the eigenvalue problem in quantum
mechanics of a particle in one-dimensional potential.
In the quantum mechanical problem  
two small energy eigenvalues can 
be realized by introducing two deep effective potential wells
isolated by a sufficiently high potential barrier. 
The two lowest energy eigenstates are given by 
superpositions of the approximate ground states in the 
respective potential wells: the lowest eigenmode becomes 
a massless mode and the other mode gains a small mass
suppressed by the tunnelling probability between the two potential wells.
The energy eigenvalues of the other modes are determined by 
the typical energy scale of the potential wells, and they are much larger 
if the potential wells are sufficiently deep or equivalently the 
barrier between two potential wells is sufficiently high. 
In this manner, the hierarchy of the energy spectrum can be realized. 
This analogy suggests that one can realize the requested mass spectrum 
in the braneworld setup, if two Randall-Sundrum type positive tension 
branes, around which the graviton is effectively 
localized owing to the effect of the bulk warp factor, 
weakly communicate with each other through a narrow throat in the bulk. 
However, the spacetime structure with such a narrow throat, 
which requires the violation of the energy condition,  
seems to be unstable in general. 
Here, the idea is to introduce 
four-dimensional Einstein-Hilbert 
terms on the branes to localize the graviton modes effectively 
near the branes, instead of considering a throat geometry, 
i.e. we adopt the five-dimensional Dvali-Gabadadze-Porrati (DGP) brane model~\cite{DGP}. 
The brane-localized Einstein-Hilbert terms 
play the role of 
the deep potential wells, and thus the two low-lying massless and 
massive graviton modes arise. 
Corresponding to the brane-localized Einstein-Hilbert terms,  
DGP two-brane model has two additional four-dimensional 
Newton's constants $\kappa_{4(\pm)}^2$, 
which effectively determine the depths of the potential wells and consequently 
the lowest KK graviton mass.
The masses of the other KK graviton modes 
are controlled by the brane separation 
and can be made large by choosing the brane separation small.
By tuning the brane separation $\dd$ to be much smaller than
$r_c^{(\pm)}:=\kappa_5^2/2\kappa_{4(\pm)}^2$, where $\kappa_5^2$ is 
the five-dimensional Newton's constant, 
we will be able to obtain the mass hierarchy 
between the lowest KK graviton mode and the other KK graviton modes. 
Then, our model would reproduce bigravity as a low energy effective 
theory.

The reproduction of bimetric theory by DGP 2-brane model 
had already been investigated in Ref.~\cite{Padilla} before dRGT bigravity was discovered.
The model in Ref.~\cite{Padilla} possesses an extra degree of freedom, 
which is called radion and absent in dRGT bigravity.
In general, two brane setup contains a low mass excitation, 
radion, corresponding to the vibration of the distance between two branes. 
Therefore, reproducing the hierarchy in the KK graviton mass spectrum is not the 
whole story.
To remove the radion,
we also introduce a stabilization mechanism of the brane separation. 
This mechanism is also necessary to keep the 
small brane separation requested for the mass hierarchy among the 
KK gravitons.  
As a concrete model of stabilization, 
we introduce a bulk scalar field with 
brane-localized potentials~\cite{stabilization}. 

It is expected that four-dimensional effective theory deduced from DGP 
two-brane model with stabilization scalar field has no BD ghost. 
Therefore, we naturally expect that bigravity derived from 
DGP two-brane model should coincide with dRGT bigravity.
However, the five-dimensional Einstein-Hilbert action in DGP two-brane model 
will introduce derivative couplings between two four-dimensional metrics 
induced on two branes.  
Thus, the correspondence between these two 
will break down if we consider higher order in gradient expansion~\cite{gradientexpansion}.
Hence, it is difficult to confirm the coincidence of the two models at the nonlinear level. 
In addition to that, in order to obtain bigravity as an effective theory of DGP 
two-brane model, we need to neglect all massive modes except for the lowest KK mode. 
This truncation is valid only when the effect of the excitation of 
massive modes are suppressed, compared with that due to the perturbation 
of our interest. 
Therefore, when we consider non-linear perturbation, 
we cannot assume that the magnitude of perturbation is infinitesimally small.   
For the reasons mentioned above, we stick to the linear perturbation 
for modes inhomogeneous in the directions parallel 
to the branes in this paper. 
The only remaining way to see the nonlinear effect will be 
changing the background energy scale, which we will discuss in this paper. 

In this paper we consider two 4-dimensional de Sitter branes and its
linear perturbation.
We will show that our model discussed above can reproduce 
bigravity effectively and that the obtained effective 
theory is identical to dRGT bigravity in the low energy regime.
Furthermore, we compare how instabilities arise in both models.  
We shall find that 
the difference in the way how instabilities develop between 
these two models breaks the correspondence in the high energy regime. 

We organize this paper as follows.
In Sec.~2 we present the setup of DGP two-brane model and its basic equations.
In Sec.~3 we study the mass spectrum in this model.
In Sec.~4 we prove that the low energy effective theory of DGP two-brane 
model is identical to dRGT bigravity.
In Sec.~5 we investigate 
how instabilities arise in these two models.
Section 6 is devoted to the summary of the paper.

\section{Model and basic equations}
Here we discuss DGP two-brane 
 model with a bulk scalar field for the
radion stabilization and give its basic equations following the
discussion in~\cite{Izumi}.
The action is given by 
\begin{align}\label{DGP2action}
   S = \frac{1}{2\kappa^2} \int dx^5 \sqrt{-g}R + \sum_{\sigma=\pm} \int
 d^4x \sqrt{ -{}^{4}\!g_{\sigma}} \left(
 \frac{1}{2\kappa_{4(\sigma)}^2}{}^{4}\!R_{(\sigma)} + L_{m\sigma}
 \right) + S_s \, , 
\end{align}
with
\begin{align}
   S_s = \int d^5x\sqrt{-g} \left( -\frac{1}{2}g^{ab}\psi_{,a}\psi_{,b} - V_B(\psi) - \sum_{\sigma=\pm}V_{(\sigma)}(\psi)\delta(y-y_{\sigma}) \right) ,
\end{align} 
where $g_{\mu\nu}$, $R$, $\!^{4}\!g_{\mu\nu}$, $\!^{4}\!R_{(\pm)}$ are 5-dimensional metric, 
5-dimensional Ricci tensor, brane-induced 4-dimensional metric and Ricci tensor, respectively.
$\kappa^2$ and $\kappa_{4(\pm)}^2$ are 
5-dimensional and 4-dimensional gravitational 
coupling constants, and $L_{m\pm}$ are the Lagrangians for the matter
fields localized
on the respective branes. 

We assume $Z_2$ symmetry across each brane. Then, 
the junction conditions imposed on the branes are derived as 
\begin{align} 
   \pm K^{(\pm)}_{\mu\nu} = r_c^{(\pm)} \left[ -\kappa_{4(\pm)}^2 \left( T_{\mu\nu}^{(\pm)} -\frac{1}{3}T^{(\pm)}g_{\mu\nu} +\frac{1}{3}V_{(\pm)}(\psi_{\pm}) \right) + \left( G_{\mu\nu}^{(\pm)} -\frac{1}{3}G^{(\pm)}g_{\mu\nu} \right)  \right]  ,
\end{align}
where $K^{(\pm)}_{\mu\nu}$, $G_{\mu\nu}^{(\pm)}$ and
$T_{\mu\nu}^{(\pm)}$ 
are the extrinsic curvatures, the induced Einstein tensors 
and the matter energy-momentum tensors on the respective brane, 
and 
\begin{align}
r_c^{(\sigma)}:=\frac{\kappa^2}{2\kappa_{4(\sigma)}^2}\, . 
\end{align}

\subsection{Background}

As the unperturbed background, we assume the bulk geometry 
\begin{align}
ds^2 = dy^2 + a^2(y)\gamma_{\mu\nu}dx^{\mu}dx^{\nu} ,
\end{align}
sandwiched by two four-dimensional de Sitter $(\pm)$-branes,  
where $\gamma_{\mu\nu}$ is four-dimensional de Sitter metric with the
comoving curvature radius $H^{-1}$. 
Then, the equations of motion become
\begin{align}\label {bg}
   \mathcal{H}' =& -\frac{\kappa^2}{3}\psi'^2 - \frac{H^2}{a^2}\, , \\
   \label{bg2}
   \mathcal{H}^2 =& \frac{\kappa^2}{6} \left(\frac{1}{2}\psi'^2-V_B \right) +  \frac{H^2}{a^2}\,  , \\
   \psi'' + &4\mathcal{H}\psi' - \frac{\partial V_B}{\partial \psi} = 0\,  ,
\end{align}
where " ${}'$ " means the partial differentiation with respect to $y$, 
and $\mathcal{H}=a'/a$.
We set the two branes at $y = y_{\pm}$ with $y_+ < y_-$.
The junction conditions on the respective branes are
\begin{align} \label{bgjc}
   \pm \mathcal{H}_{\pm} = r_c^{(\pm)} {H^2\over a^2}-
 \frac{\kappa^2}{6}V_{(\pm)}\left( \psi_{\pm} \right) \, ,
\end{align}
and
\begin{align} 
   \psi' _{\pm} =\pm 
  \frac{1}{2}\left.\frac{\partial V_{(\pm)}}{\partial \psi}\right\vert_{y=y_{\pm}}
\, .
\end{align}

\subsection{Perturbation}

Now we consider perturbation around the background mentioned above.
We use Newton gauge, in which the spin-0 components of the shear of the
hypersurface normal vector and the shift vector are set to zero.
In this gauge, using traceless part and $\{y\mu\}$-components of the Einstein
equations, we find that perturbation of the metric, $h_{ab}$, and that
of the scalar field, $\delta \psi$, are written as 
\begin{align}
&   h_{yy} = 2\phi\, , \\
&   h_{y\mu} = 0\, ,\\
&   h_{\mu\nu} = h_{\mu\nu}^{(TT)} - \phi \tilde{\gamma}_{\mu\nu}\, , \\
&\delta \psi = \frac{3}{2\kappa^2\psi'} [\partial_y +2\mathcal{H}] \phi\, ,
\end{align}
where  $\tilde{\gamma}_{\mu\nu}:=a^2(y)\gamma_{\mu\nu}$.
The bulk equations for $h_{\mu\nu}^{(TT)}$ become
\begin{align}\label{TTb}
&   \left[ \hat{L}^{(TT)} + {1\over a^{2}} \left({}^4\Box - 2H^2
\right)\right] h_{\mu\nu}^{(TT)} = 0\, , 
\end{align}
with
\begin{align}
&   \hat{L}^{(TT)} := \frac{1}{a^2} \partial_y a^4 \partial_y\frac{1}{a^2}\, ,
\end{align}
where ${}^4\Box:=\gamma^{\mu\nu}\nabla_{\mu}\nabla_{\nu}$, and
$\nabla_{\mu}$ is the covariant differentiation associated with
$\tilde{\gamma}_{\mu\nu}$.
In raising or lowering Greek indices, we use $\tilde{\gamma}_{\mu\nu}$.
The bulk equation for the scalar-type perturbation becomes
\begin{align}\label{sb}
&   \left[ \hat{L}^{(\phi)} + \frac{ {}^4\Box + 4H^2 }{\psi'^2} \right] \phi = 0\, , 
\end{align}
with
\begin{align}
&   \hat{L}^{(\phi)} := a^2 \partial_y \frac{1}{a^2\psi'^2} \partial_y a^2 - \frac{2\kappa^2}{3}a^2\, .
\end{align}
In order to derive the junction conditions, it is
convenient to use the
Gaussian normal coordinates, in which the lapse function 
and the shift vector are set
to 1 and ${\bf 0}$, respectively, and the brane locations are not perturbed. 
Here we discriminate the variables in the Gaussian normal coordinates by 
associating a bar like $\bar{h}_{\mu\nu}^{(\pm)}$.
Since 
the Gaussian normal coordinates with respect to the $(+)$-brane
are in general different from those with respect to $(-)$-brane, 
we also associate the subscript $(\pm)$ to distinguish them.
In these coordinates, 
the junction conditions for metric perturbation are given by 
\begin{align} 
   \pm(\partial_y - 2\mathcal{H})\bar{h}_{\mu\nu}^{(\pm)} 
=& -\kappa^2\left[ T_{\mu\nu}^{(\pm)} -\frac{1}{3}T^{(\pm)}
 \tilde{\gamma}_{\mu\nu} \right] \mp \frac{2\kappa^2}{3}
 \tilde{\gamma}_{\mu\nu}\psi'\delta\psi + 2r_c^{(\pm)} \left[
 X_{\mu\nu}^{(\pm)} -\frac{1}{3} X^{(\pm)} \tilde{\gamma}_{\mu\nu}
 \right]\, , 
\end{align}
with
\begin{align}
   X_{\mu\nu}^{(\pm)} := & -\frac{1}{2} \left( {1\over
 a^2}{}^4\Box\bar{h}_{\mu\nu}^{(\pm)} -
 \nabla_{\mu}\nabla_{\alpha}\bar{h}^{(\pm)\alpha}_{\nu} 
- \nabla_{\nu}\nabla_{\alpha}\bar{h}^{(\pm)\alpha}_{\mu} + \nabla_{\mu}
 \nabla_{\nu} \bar{h}^{(\pm)} \right) \nonumber \\
& \ \ \  -\frac{1}{2}\tilde \gamma_{\mu\nu} \left( \nabla_{\alpha}
 \nabla_{\beta}\bar{h}_{(\pm)}^{\alpha\beta} -{1\over
 a^{2}}{}^4\Box\bar{h}^{(\pm)} \right)
+{H^2\over a^{2}} \left( \bar{h}_{\mu\nu}^{(\pm)} + \frac{1}{2}
 \tilde{\gamma}_{\mu\nu} \bar{h}^{(\pm)} \right)\, .
\label{eq:defX}
\end{align}
The junction conditions for the bulk scalar field become
\begin{align} 
   \pm 2\delta \bar{\psi}' = V''^{(\pm)}(\psi) \delta \bar{\psi} \, .
\end{align}
The generators of the gauge transformation from the Gaussian normal
coordinates to the Newton gauge are 
\begin{align} 
   \xi^y_{(\pm)} =& \int^y_{y_{\pm}} \phi(y')dy' + \hat{\xi}^y_{(\pm)}(x^{\mu})\, , \\
   \xi^{\nu}_{(\pm)} =& - \int^y_{y_{\pm}} \tilde{\gamma}^{\mu\nu}(y')\left[ \int^{y'}_{y_{\pm}} \phi_{,\mu}(y'')dy'' + \hat{\xi}^y_{(\pm) ,\mu}(x^{\rho}) \right] dy' + \hat{\xi}^{\nu}_{(\pm)}(x^{\rho})\, ,
\end{align}
where $\hat{\xi}^y_{(\pm)} (x^{\mu}) = y|_{brane} - y_{\pm}$ represents
the perturbed brane position in the coordinates of Newton gauge, which 
we simply call the brane bending. 
Under this gauge transformation, 
the perturbation variables in two gauges transform as 
\begin{align} 
   \bar{h}^{(\pm)}_{\mu\nu} =& h_{\mu\nu} - 2\nabla_{(\mu} \xi_{\nu)}^{(\pm)} - 2\mathcal{H}\tilde{\gamma}_{\mu\nu} \xi^y_{(\pm)}\, , \\
   \delta \bar{\psi}^{\pm} =& \delta \psi^{\pm} - \psi' \xi^y_{(\pm)}\, .
\end{align}
From these relations, we obtain the junction conditions in Newton gauge.
The conditions for the traceless part of metric perturbation become
\begin{align} 
   \pm(\partial_y - 2\mathcal{H}_{\pm}) h_{\mu\nu}^{(TT)} = -\kappa^2
 \Sigma_{\mu\nu}^{(\pm)} - r_c^{(\pm)} a_{\pm}^{-2} ({}^4\Box - 2H^2)
 h_{\mu\nu}^{(TT)}\, ,
\end{align}
where
\begin{align} 
   \Sigma^{(\pm)}_{\mu\nu} := \left( T_{\mu\nu}^{(\pm)}
 -\frac{1}{4}T^{(\pm)} \tilde{\gamma}_{\mu\nu}^{\pm} \right) \pm
 \frac{2}{\kappa^2} 
   \left( \nabla_{\mu}\nabla_{\nu} -\frac{1}{4} \gamma_{\mu\nu} {}^4\Box
 \right) Z_{(\pm)}\, , 
\end{align}
with
\begin{align}
   \label{Z}
   Z_{(\pm)} := (1 \mp 2r_c^{(\pm)} \mathcal{H}_{\pm}) \hat{\xi}^y_{\pm}
 \mp r_c^{(\pm)} \phi (y_{\pm})\, .
\end{align}
The condition for the trace part of metric perturbation leads
\begin{align} 
   a^{-2}_{\pm} ({}^4\Box + 4H^2) Z_{(\pm)} = \pm \frac{\kappa^2}{6}
 T^{(\pm)}\, .
\end{align}
The condition for the scalar-field perturbation becomes
\begin{align} \label{jcs}
   \mp \frac{2\kappa^2}{3} \left( \delta \psi - \psi'(1\mp2r_c^{(\pm)}
 \mathcal{H})^{-1}(Z_{(\pm)} \pm r_c^{(\pm)} \phi) \right) =
 \frac{\epsilon^{(\pm)}}{a^2 \psi'} ( {}^4\Box + 4H^2) \phi\, ,
\end{align}
where
\begin{align} 
   \epsilon^{(\pm)} := \frac{2}{V''^{(\pm)} \mp 2\psi''/\psi'}\, .
\end{align}
Finally, combining the bulk equations and junction conditions for the tensor-type perturbation, we find
\begin{align} \label{TT}
   \left[ \hat{L}^{(TT)} + {{}^4\Box - 2H^2\over a^2} \right]
 h_{\mu\nu}^{(TT)} = \sum_{\sigma=\pm}\left( -2\kappa^2
 \Sigma^{(\sigma)}_{\mu\nu} -2r_c^{(\sigma)} a_{\sigma}^{-2} ({}^4\Box -
 2H^2) h_{\mu\nu}^{(TT)} \right) \delta(y-y_{\sigma})\, .
\end{align}
For the scalar-type perturbation, we find
\begin{align} \label{s}
   \left[ \hat{L}^{(\phi)} + \frac{{}^4\Box +4H^2}{\psi'^2} \right]
 \phi = \sum_{\sigma=\pm}\left(\frac{ 4a^2\kappa^2}{3 
(\sigma 1 -2 r_c^{(\sigma)} \mathcal{H}_{\sigma})} 
(Z_{(\sigma)} +\sigma
 r_c^{(\sigma)} \phi) - \frac{2\epsilon^{(\sigma)}} {\psi'^2} ({}^4\Box
 + 4H^2) \phi \right) \delta(y-y_{\sigma})\, .
\end{align}

\section{Mass spectrum}

In this section we show that 
%
it is possible to make the mass hierarchy among KK gravitons, 
and we obtain bigravity as the low-energy effective theory of our
model by properly introducing the stabilization mechanism.
For simplicity, we set $y_+ = 0$, $y_- = \dd$, $a_+ = 1$.

\subsection{eigenvalue problems that determine the mass spectrum}
To see the mass spectrum, we set the source terms 
$\Sigma_{\mu\nu}^{(\pm)}$ and $Z_{(\pm)}$ to zero and separate the
variables in Eq.~\eqref{TT} and Eq.~\eqref{s}.
From Eq.~\eqref{TT}, we define an eigenvalue problem 
\begin{align}
  \left[ \frac{m_i^2}{a^2} \left( 1 + 2\sum_{\sigma=\pm}r_c^{(\sigma)}
 \delta(y-y_{\sigma}) \right) \right] u_i(y) = -\hat{L}^{(TT)}u_i (y)\, , \
\end{align}
where the operator ${}^4\Box - 2H^2$ was replaced with 
the eigenvalues $m_i^2$, and $u_i (y)$ are the corresponding
eigenfunctions. 
Also, we define the inner product
\begin{align}\label{inprou} 
   ( u_i , {} u_j )^{(TT)} := \oint \frac{dy}{a^2} \left( 1 +
 2\sum_{\sigma = \pm} r_c^{(\sigma)} \delta (y-y_{\sigma}) \right)
 u_i(y) u_j(y) =\delta_{ij}\, ,
\end{align}
with respect to which 
eigenmodes with different eigenvalues are mutually orthogonal.
By definition, the norm $(u_i, {} u_i)^{(TT)}$ is always positive.
Using these mode functions $u_i(y)$, we can also obtain the solution for
Eq.~\eqref{TT} with the source term as
\begin{align}\label{solTT}
   h_{\mu\nu}^{(TT)}(y) = -2\kappa^2 \sum_{i} \frac{u_i (y_+) u_i (y) }{
 {}^4\Box - 2H^2 - m_i^2 } \Sigma_{\mu\nu}^{(+)}\, .
\end{align}
Here we assumed that only the $(+)$-brane has the source 
$\Sigma_{\mu\nu}^{(+)}$.

Similarly, from Eq.~\eqref{s}, we define 
an eigenvalue problem 
\begin{align} \label{v}
  \frac{ \mu_i^2 + 4H^2 }{\psi'^2} \left( 1 + \sum_{\sigma = \pm} 2 \epsilon^{(\sigma)} \delta (y-y_{\sigma}) \right) v_i (y) 
  = \left[ -\hat{L}^{(\phi)} + \sum_{\sigma=\pm} \frac{4r_c^{(\sigma)}
 \kappa^2 a^2}{3 (1 - \sigma 2r_c^{(\sigma)} \mathcal{H}_{\sigma} )}
 \delta(y-y_{\sigma}) \right] v_i (y)\,,
\end{align}
where the operator ${}^4\Box$ was replaced with the eigenvalues $\mu_i^2$, 
and $v_i(y)$ are the corresponding eigenfunctions. 
Also, we define the inner product
\begin{align} 
   ( v_i , {} v_j)^{(\phi)} := \oint \frac{dy}{\psi'^2} \left( 1 +
 \sum_{\sigma = \pm} 2 \epsilon^{(\pm)} \delta (y-y_{\sigma}) \right)
 v_i(y) v_j(y) =\delta_{ij}\, .
\end{align}
We assume $\epsilon^{(\pm)}>0$ to guarantee that the inner product of
$v_i$ is positive definite.  
This assumption is easily satisfied when both $V''_{(+)}$ 
and $V''_{(-)}$ are sufficiently large positive.
Using $v_i(y)$, we can also find the solution for Eq.~\eqref{s} 
with the source term as
\begin{align}\label{sols}
   \phi(y) = \frac{4\kappa^2 a_+^2}{3} (1 - 2r_c\mathcal{H}_+)^{-1}
 \sum_{i} \frac{v_i (y_+) v_i (y) }{ {}^4\Box - \mu_i^2 } Z_{(+)}\, .
\end{align}
Again, we assumed that only the $(+)$-brane has the source $Z_{(+)}$.

\subsection{Tensor-type perturbation modes}
\label{sec:tensor}
We begin the detailed analysis with the tensor-type perturbation. 
The bulk equation for the eigenfunctions $u_i$ can be written as 
\begin{align}\label{mTTb}
   \hat{L}^{(TT)} u_i = - a^{-2} m_i^2  u_i\, .
\end{align}
The junction conditions are
\begin{align}\label{mTTjc}
   \pm(\partial_y - 2\mathcal{H}_{\pm}) u_i = - r_c^{(\pm)} a^{-2} m_i^2 u_i\, .
\end{align}
Using these equations, we find that $u_0 = C_0 a^2$ is a massless mode, 
where $C_0$ is a constant such that properly 
normalizes the mode with respect to the inner product \eqref{inprou}. 
As we are interested in low mass modes,  
we assume that the mass eigenvalue
of the first KK graviton mode 
$m_1$ is small enough to satisfy $m_1\dd\ll1$.
Here, we assume that $\mathcal{H}$ is sufficiently small and five-dimensional
scale factor $a(y)$ does not largely deviate from unity.
These assumptions are required in order to avoid scalar instability, which is shown
later in Sec.~\ref{scalarmode}.
Introducing the new non-dimensional coordinate $Y=y/\dd$, 
we rewrite the equations \eqref{mTTb} and \eqref{mTTjc} as
\begin{align}\label{mTTb2}
   \frac{1}{a^2} \partial_Y a^4 \partial_Y \frac{1}{a^2} u_1 = -
 \frac{(m_1\dd)^2}{a^2}  u_1\, ,
\end{align}
and
\begin{align}\label{mTTjc2}
   \pm \left( \partial_Y - 2 \frac{\partial_Y a}{a} \right) u_1 = - \frac{m_1^2 r_c^{(\pm)} \dd} {a^2} u_1\, .
\end{align}
We restrict our attention to the mass range 
$ 
m_1^2 \ll \dd^{-2}$.  
Hence, neglecting the r.h.s.~of Eq.~\eqref{mTTb2}, 
we find an approximate solution of the bulk equation as 
\begin{align}\label{u1}
   u_1^{(0)} \propto a^2 \int^Y a^{-4}dY'\, .
\end{align}
The junction conditions determine the mass eigenvalue and the
integration constant, which are 
yet undetermined in the expression~\eqref{u1}, simultaneously.
The mass eigenvalue is given by
\begin{align}\label{TTmass1}
   m_1^2 = \frac{1}{\dd \int^{Y_-}_{Y_+}a^{-4}dY'}
 \left(\frac{1}{a_+^2r_c^{(+)}}+\frac{1}{a_-^2r_c^{(-)}}\right)\, ,
\end{align}
while the mode function at the leading order becomes
\begin{align} \label{TTfcn}
   u_1^{(0)} = C_1 a^2\left( 1 - r_c^{(+)} \dd\, m_1^2 \int^{Y}_{0}a^{-4}dY
 \right)\, ,
\end{align}
where $C_1$ is the normalization constant.
We find that $m_1^2$ is the unique eigenvalue that satisfies $m_1^2
\ll \dd^{-2}$.
The mode function \eqref{TTfcn} has only one node, 
which is consistent with the fact
that this mode is the first KK graviton mode.  
Since the other KK graviton modes have mass at least comparable 
to $\dd^{-2}$, we find that the mass hierarchy between $m_1^2$ and
$m_{i\geq 2}^2$ is realized when $r_c^{(\pm)} \gg \dd$, i.e. $m_1^2
\simeq (r_c^{(\pm)}\dd)^{-1} \ll \dd^{-2} \simeq m_2^2$.

\subsection{Scalar-type perturbation modes}
\label{scalarmode}
We estimate the lowest eigenvalue of the scalar mode. 
Here, we set $H\simeq0$, for simplicity. 
In the absence of the stabilization scalar field, there should be a massless 
degree of freedom corresponding to the fluctuation of brane separation.
In fact, Eq. \eqref{v} has a zero eigenvalue mode and the corresponding eigenfunction 
$v_0 \propto a^{-2} $ when $\psi' = 0$.
Therefore, if we assume that the back reaction
of the stabilization scalar field 
to the background geometry is weak, i.e. 
\begin{align}
     \frac{| \mathcal{H}' |}{ \mathcal{H}^2 } = \frac{\kappa^2 \psi'^2
  }{ 3\mathcal{H}^2 } \ll 1\, ,
\end{align}
we can perturbatively obtain a small mass eigenvalue. 
In Eq.~\eqref{v} we can treat the terms that are not enhanced by a 
factor $1/\psi'{}^2$ in the square brackets on the
r.h.s.~as perturbation under this weak back reaction approximation.
Then, we obtain the leading order correction to the almost zero mode 
eigenvalue as
\begin{equation} \label{smass}
   \mu^2 \approx  
   \frac{\displaystyle 2 \int^{y_-}_{y_+} \frac{dy}{\displaystyle a^2} 
   + \sum_{\sigma} \frac{\displaystyle 2 r_c^{(\sigma)} }{\displaystyle a_{\sigma}^2 } 
   \frac{1}{1 - \sigma 2r_c^{(\sigma)} \mathcal{H}_{\sigma} }  } {
 \displaystyle \int^{y_-}_{y_+} \frac{dy}{a^4(-\mathcal{H}' ) } + \displaystyle \sum_{\sigma} 
 \frac{ \epsilon^{(\sigma)} }{a_{\sigma}^4(-\mathcal{H}'_{\sigma} ) }
 }\, .
\end{equation}
In the following discussion we set $\epsilon^{(\pm)}$ to zero for
simplicity.
Using weak back reaction approximation, we can set 
$\mathcal{H} \approx$constant 
and $a \approx \mathrm{e}^{\mathcal{H}y}$ in Eq.~\eqref{smass}.
Then, the above correction to $\mu^2$ 
is reduced to 
\begin{equation} 
\label{smass2}
   \mu^2 \approx 
  \left.    {1\over \mathcal{H}}\left[ 
  \frac{1}{1 - 2r_c^{(+)} \mathcal{H} }  
  -  \frac{e^{-2\mathcal{H}\dd}}{1 + 2r_c^{(-)} \mathcal{H} }  
\right]\right/
 \displaystyle \int^{y_-}_{y_+} \frac{dy}{a^4(-\mathcal{H}')}\, ,
\end{equation}
and turns out to be positive as long as 
$1+2 r_c^{(-)} \mathcal{H}_->0$ 
is satisfied for negative ${\mathcal{H}}$.
For positive ${\mathcal{H}}$, the analogous condition for $\mu^2$ 
to be positive is 
$1-2 r_c^{(+)} \mathcal{H}_+>0$. 
The present approximation for $\mu^2$ is necessarily invalid
when the value of $1\mp 2 r_c^{(\pm)}\mathcal{H}_{\pm}$ crosses zero. 
After crossing the critical value, the above final expression 
stays negative. 

The meaning of the critical value $1\mp 2
r_c^{(\pm)}\mathcal{H}_{\pm}=0$ can be interpreted as follows.
Combining the background bulk equation~\eqref{bg2} evaluated at $y=y_{\pm}$
and the junction conditions~\eqref{bgjc}, we find quadratic
equations for $\mathcal{H}_{\pm}$ as
\begin{align}
\mathcal{H}_{\pm}^2 \pm \frac{1}{r_c^{(\pm)} } \mathcal{H}_{\pm} +
 \frac{\kappa^2}{6} \left( -\frac{1}{2} \psi_{\pm}'^2 +V_B(\psi_{\pm})
 +\frac{1}{r_c^{(\pm)}} V_{(\pm)}(\psi_{\pm}) \right) = 0\, .
\end{align}
Then, we obtain the solutions for $\mathcal{H}_{\pm}$ as 
\begin{align}\label{branch+}
2r_c^{(+)} \mathcal{H}_{+} -1 = \pm \sqrt{1-\frac{2}{3} \kappa^2
 r_c^{(+)} \bar{V}_{(+)} }\, ,
\end{align}
at $y=y_+$ and 
\begin{align}\label{branch-}
2r_c^{(-)} \mathcal{H}_{-} +1 = \pm \sqrt{1-\frac{2}{3} \kappa^2 r_c^{(-)} \bar{V}_{(-)} }\, ,
\end{align}
at $y=y_-$, where $\bar{V}_{(\pm)} := -\frac{1}{2} \psi_{\pm}'^2
+V_B(\psi_{\pm}) +\frac{1}{r_c^{(\pm)}} V_{(\pm)}(\psi_{\pm}) $.
For $1\mp 2 r_c^{(\pm)}\mathcal{H}_{\pm} \neq 0$, two 
solutions of $\mathcal{H}_{\pm}$ exist for the same value of $\bar{V}_{(\pm)}$ and they correspond to the normal and
self-accelerating branches. We can choose the normal or
self-accelerating branch by the choice of appropriate signs on the r.h.s.~of
Eqs.~\eqref{branch+} and \eqref{branch-}. 
When the condition $1\mp 2 r_c^{(\pm)}\mathcal{H}_{\pm}=0$ is 
satisfied, however, the two
branches degenerate, and hence this condition defines the boundary of 
the two branches. 
According to Eq.~\eqref{Z}, we can also understand that 
the sign of the brane bending $\xi^y_{(\pm)}$ becomes indefinite 
at the critical point, $1\mp2r_c^{(\pm)}\mathcal{H}_{\pm}=0$. 
We define that the solution is in the normal branch when the conditions 
$1\mp2r_c^{(\pm)} \mathcal{H}_{\pm} >0$ are satisfied. 
Later, we will show that there arises a tachyonic scalar mode before 
either of $1\mp2 r_c^{(\pm)} \mathcal{H}_{\pm}$ changes its signature. 
Namely, the self-acceleration branch is unstable. 
In order to avoid this tachyonic instability, 
$|\mathcal{H}|$ should be kept small, 
which is consistent with the conditions 
used in deriving the estimate of $m_1^2$ in Sec.~\ref{sec:tensor}. 

We should make the lowest mass of the scalar modes much larger than that
of the lowest KK graviton mode to reproduce bigravity as 
the low energy effective theory. 
As we can see from the above expression for $\mu^2$, this mass eigenvalue
can be made large keeping $r_c^{(\pm)} \mathcal{H}_{\pm}$ small, if 
$|\mathcal{H}' |$ is sufficiently large.
In the numerator of r.h.s. of Eq.~\eqref{smass}, the first term is much less 
than other terms and can be ignored when we assume $\dd / r_c^{(\pm)} \ll 1$ 
to realize the hierarchy among KK gravitons.
Therefore, we would be able to make $|\mathcal{H}'|$ 
as large as $1/\dd r_c^{(\pm)}$ without 
violating the condition $|r_c^{(\pm)} \mathcal{H}|\lesssim 1$
and expect $\mu^2$ can be made as large as $O(\dd^{-2})$.  
However, the parameter range in which 
$|\mathcal{H}' \dd r_c^{(\pm)}| \gg 1$ is outside the validity range of 
the above perturbative derivation of the expression for $\mu^2$.  
Therefore, we show it numerically that the models that realize the 
requested hierarchy really exist in the succeeding subsection.   

\subsection{Numerical proof of the existence of models that realize hierarchy}

\begin{figure}[H]
  \begin{center}
   \includegraphics[width=100mm]{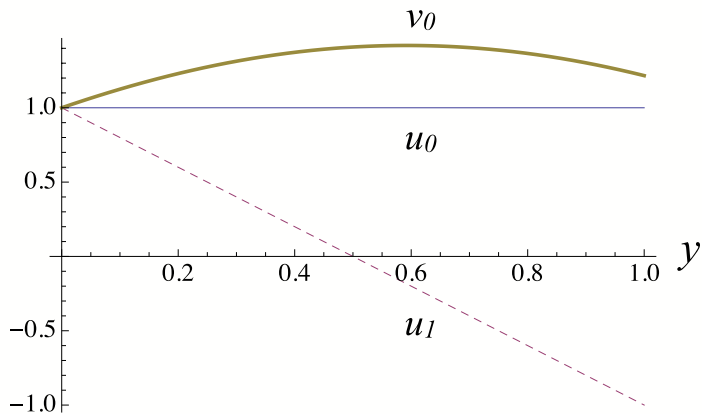}
  \end{center}
  \caption{The mode functions $u_0(y)$, $u_1(y)$ and $v_0(y)$.
  The solid, dotted and thick lines represent $u_0(y)$, $u_1(y)$ and $v_0(y)$ respectively.}
  \label{fig1}
\end{figure}

To show it possible to realize the requested mass hierarchy, we
numerically solve the above eigenvalue problems.
Here, we consider two Minkowski branes, setting $H$ to zero. 
We construct an explicit solution of the background geometry and the
stabilization scalar field by choosing the scalar-field potential,
following Ref.~\cite{deWolfe}, as
\begin{align} 
  V(\psi) = \frac{1}{8} \left( \frac{\partial W(\psi)}{\partial \psi} \right)^2 - \frac{\kappa^2}{6} W(\psi)^2 \, .
\end{align}
Adopting this potential form, the equations for $\mathcal{H}$ and $\psi$ are decoupled as
\begin{align} 
   \psi' = \frac{1}{2} \frac{\partial W(\psi)}{\partial \psi} \, , \\
   \mathcal{H} = -\frac{\kappa^2}{6} W(\psi) \, .
\end{align}
As a simple example, we choose the form of the bulk potential $W(\psi)$ and the brane-localized potentials $V_{\pm}$ as
\begin{align} 
   W(\psi) &= \frac{3}{L} - b \psi^2\, ,   \\
   V_{(+)} (\psi) &= W(\psi_+) + W' (\psi_+)(\psi - \psi_+) + \gamma_{(+)} (\psi - \psi_+)^2\, , \\
   V_{(+)} (\psi) &= - W(\psi_-) - W' (\psi_-)(\psi - \psi_-) + \gamma_{(-)} (\psi - \psi_-)^2\, ,
\end{align}
where $L$, $b$ and $\gamma_{(\pm)}$ are model parameters. To make $\epsilon^{(\pm)}$ small, we should take $\gamma_{(\pm)}$ large. 
For these potentials, we can analytically obtain the solution for $\psi$ and $\mathcal{H}$ as
\begin{align} 
   \psi &= \psi_0 \mathrm{e}^{-by}\, , \\
   \mathcal{H} &= -\frac{\kappa^2}{6} \left( \frac{3}{L} - b \psi_0^2 \mathrm{e}^{-2by} \right)\, .
\end{align}
When $b\dd \ll 1$ and $\kappa \psi_0 \lesssim 1$, $\left| \mathcal{H}' / \mathcal{H}^2 \right|$ can be large in the whole spacetime by tuning $L$.
On the other hand, the conditions $1\mp2 r_c^{(\pm)}
\mathcal{H}_{\pm}>0$, which guarantee the positivity of the lowest
scalar-mode mass squared $\mu^2_0$, imply the condition $(\kappa \psi_0
b\dd)^2 \lesssim \dd/r_c$ whose r.h.s.~should be much less than unity
when we request the hierarchy among KK graviton masses.
Therefore, we expect that the lowest scalar-mode mass has a large
positive value when we set $\kappa \psi_0 b\dd$ very small and tune $L$,
and at the same time we realize the mass hierarchy among KK gravitons.
Then, we numerically confirm that we can realize the requested mass
hierarchy between the lowest KK graviton mode and the other
massive modes. 
We present the result of the numerical calculation where we set the
parameters to be $\kappa^2=1.00$, $\dd=1.00$, $r_c^{(\pm)}=1.00\times
10^5$, $L=3.00\times 10^3$, $b=1.00\times 10^{-2}$,
$\gamma_{(\pm)}=1.00\times 10^3$ and $\psi_0 = 0.318$.
Figure~\ref{fig1} shows the mode functions $u_0(y)$, $u_1(y)$ and
$v_0(y)$.
The mass eigenvalues are obtained as $m_1^2=2.00\times
10^{-5}$, $m_2^2=9.87$ and $\mu_0^2=1.77$.
From the above calculation, we find it possible to construct a higher
dimensional model whose low energy effective theory has only two
gravitons: one is massless and the other has a tiny mass.

\section{linear perturbation in $\rm\bf d$RGT bigravity and DGP two-brane model}

In the preceding section, we concluded that we can realize bigravity
as the low energy effective theory of DGP two-brane model 
with an appropriate stabilization mechanism. 
This effective theory has 7 degrees of freedom in the gravity sector.  
Whilst, dRGT bigravity, which is constructed 
for Boulware-Deser not to appear, also has healthy 7 degrees of freedom. 
If dRGT bigravity is the unique ghost-free bigravity theory, 
the effective theory of DGP two-brane model should 
coincide with dRGT bigravity when all massive modes in DGP two-brane
model except for the lowest KK graviton are sufficiently heavy and decouple. 
However, the action of DGP two-brane model \eqref{DGP2action}, 
is not only composed of Ricci scalars with respect to 
two metrics induced on the respective branes 
but also contains five-dimensional Ricci scalar in the bulk, 
whose counterpart seems to be absent in dRGT bigravity. 
Hence, if we consider higher order in the derivative expansion, 
the two metrics will have derivative coupling in DGP two-brane model, 
and the correspondence will not be maintained. 
Also, in order to obtain bigravity as an effective theory of DGP two-brane model, 
all massive modes except for the lowest KK mode must be suppressed. 
Thus, when we consider non-linear perturbation, it would be necessary 
to consider relatively large magnitude of perturbation. 
Otherwise, the effect of heavy modes will be larger than the 
non-linear effect. Hence, it is not clear if we can extend 
our analysis to non-linear level. 
For these reasons, instead of pursuing 
the extension of our analysis to the non-linear perturbation, here 
we investigate the nonlinear effect just by changing the background energy scale.
Below, we consider perturbation around de Sitter brane background 
with arbitrary energy scale $H$.  
We will show that two models are identical 
as long as the linear perturbation is concerned
and how the parameters in two models correspond.

\subsection{DGP model}

Considering perturbation from the de Sitter brane background 
caused by the matter on the $(+)$-brane
in DGP model, the metrics induced on the branes become
\begin{align} \label{lDGP}
   \bar h_{\mu\nu} (y_{\pm}) + \nabla_{\mu} \hat{\xi}_{\nu} + \nabla_{\nu}
 \hat{\xi}_{\mu} = h_{\mu\nu}^{(TT)}(y_{\pm}) - \tilde{\gamma}_{\mu\nu}
 \left( \phi(y_{\pm}) + 2\mathcal{H}_{\pm} \hat{\xi}^y_{\pm} \right)\, .
\end{align}
Following the discussion in Ref.~\cite{Izumi}, from Eqs.~\eqref{solTT} and 
\eqref{sols}, the metric components induced on the respective 
branes are obtained as
\begin{align}
   h^{(TT)}_{\mu\nu}(y_{\pm}) =& -2\kappa^2 \sum_i
 \frac{u_i(y_+)u_i(y_{\pm})}{{}^4\Box -2H^2 - m_i^2} \left[
 T_{\mu\nu}^{(+)} -\frac{1}{4}\tilde{\gamma}_{\mu\nu}T^{(+)} +
 \frac{a_+^2}{3(m_i^2 - 2H^2)} \left( \nabla_{\mu}\nabla_{\nu}
 -\frac{1}{4} \gamma_{\mu\nu} {}^4\Box \right)T^{(+)} \right] 
 \nonumber \\
   &\qquad\qquad\qquad
    +\frac{2\kappa^2a_+^2}{3} \left( \frac{u_i(y_+)u_i(y_{\pm})}{m_i^2 -
 2H^2} \right) \left( \nabla_{\mu}\nabla_{\nu} -\frac{1}{4}
 \gamma_{\mu\nu} {}^4\Box \right) \frac{1}{{}^4\Box + 4H^2} T^{(+)}\, , \\
   \phi_{+} + 2\mathcal{H}_{+} \hat{\xi}^y_{+} &= \frac{2\kappa^4
 a_+^4}{9\left(2r_c^{(+)}\mathcal{H}_+ -1 \right)^2} \sum_i
 \frac{v_i(y_+)^2}{\mu_i^2 + 4H^2} \frac{1}{{}^4\Box - \mu_i^2} T^{(+)}
 \nonumber \\
      &\qquad
 - \frac{\kappa^2 a_+^2}{3\left(2r_c^{(+)}\mathcal{H}_+ -1\right)}
 \left[
 \frac{2\kappa^2 a_+^2}{3\left(2r_c^{(+)}\mathcal{H}_+ -1 \right)} \left(
 \sum_i \frac{v_i(y_+)^2}{\mu_i^2 + 4H^2} \right) + \mathcal{H}_{\pm}
 \right]  \frac{1}{{}^4\Box + 4H^2} T^{(+)}\, , \\
   \label{phiy-}
   \phi_{-} + 2\mathcal{H}_{-} \hat{\xi}^y_{-} &= \frac{2\kappa^4
 a_+^4}{ 9\left(2r_c^{(+)}\mathcal{H}_+ -1 \right)
     \left(2r_c^{(-)}\mathcal{H}_- +1 \right)}
 \sum_i \frac{v_i(y_+)v_i(y_-)}{\mu_i^2 + 4H^2}  \left[
 \frac{1}{{}^4\Box - \mu_i^2} T^{(+)} -  \frac{1}{{}^4\Box + 4H^2}
 T^{(+)} \right]\, .
\end{align} 
Since there is no propagating degree of freedom corresponding to 
$\mu^2=-4H^2$, the terms in $h_{\mu\nu}(y_{\pm})$ proportional to
$({}^4\Box+4H^2)^{-1}T^{(+)}$ should not be present in total. 
This condition implies an identity
\begin{align} \label{id}
   2 \left( \sum_i \frac{u_i^2(y_+)}{m_i^2 - 2H^2} \right) +
 \frac{a_+^2}{H^2\left(2 r_c^{(+)} \mathcal{H} _+- 1\right)} \left[
 \frac{2\kappa^2a_+^2}{3\left(2 r_c^{(+)} \mathcal{H} _+- 1\right)} \left( \sum_i
 \frac{v_i^2(y_+)}{\mu_i^2 + 4H^2} \right) + \mathcal{H}_+ \right] = 0\,
 ,
\end{align}
which can be also proven by using the residual gauge transformation and
junction condition on the $(+)$-brane~\cite{Izumi}. 
Applying the same argument to the junction condition on the $(-)$-brane, 
we obtain another identity:
\begin{align} \label{id-}
   \left( \sum_i \frac{u_i(y_+)u_i(y_-)}{m_i^2 - 2H^2} \right) +
 \frac{\kappa^2
 a_+^2 a_-^2}{3 H^2\left(2 r_c^{(+)} \mathcal{H} _+- 1\right) \left(2 r_c^{(-)} \mathcal{H} _- + 1\right)} \left( \sum_i
 \frac{v_i(y_+)v_i(y_-)}{\mu_i^2 + 4H^2} \right) = 0\, .
\end{align} 
Using these relations the expressions for the induced metrics simplify 
a lot. Furthermore, for the comparison with the bigravity model, 
we truncate the expression by keeping only the 
massless and the lowest mass KK graviton modes. 
Eliminating the terms that can be erased by 
four-dimensional gauge transformation, 
$\bar h_{\mu\nu}(y_{\pm})$ become
\begin{align} 
   \bar h_{\mu\nu} (y_{\pm}) = -2\kappa^2 \sum_{i=0, 1}
 \frac{u_i(y_+)u_i(y_{\pm})}{{}^4\Box -2H^2 - m_i^2} \left(
 T_{\mu\nu}^{(+)} - \left( 1 + \frac{1}{3} \frac{{}^4\Box -2H^2 -m_i^2}
 {{}^4\Box +6H^2 -m_i^2} \frac{m_i^2 - 6H^2}{m_i^2 -
 2H^2}  \right)\frac{1}{4} \tilde{\gamma}_{\mu\nu}T^{(+)} \right)\, .
\end{align}
We can confirm that the massless mode behaves like the graviton in 
GR, while the massive mode behaves like the 
Fierz-Pauli massive graviton. 

We can write the metrics induced on the branes
as 
\begin{align} \label{DGP}
   \bar h_{\mu\nu} (y_{\pm})&= 2r_c^{(+)} h^{(0)}_{\mu\nu} u_0(y_+)u_0(y_{\pm}) + 2r_c^{(+)}h^{(1)}_{\mu\nu} u_1(y_+)u_1(y_{\pm})\, , \\
   \label{graviton}
   \left(\Box -2H^2 \right)h_{\mu\nu}^{(0)} &= -2\kappa_{4(+)}^2 
 \left( T_{\mu\nu}^{(+)} -  \left( 1 + \frac{{}^4\Box -2H^2}
 {{}^4\Box +6H^2 } \right) \frac{1}{4}\tilde{\gamma}_{\mu\nu}T^{(+)} \right)\, , \\
   \label{mgraviton}
   (\Box -2H^2 -m_1^2) h_{\mu\nu}^{(1)} &= -2\kappa_{4(+)}^2
  \left( T_{\mu\nu}^{(+)} -  \left( 1 + \frac{1}{3} \frac{{}^4\Box -2H^2 -m_1^2}
 {{}^4\Box +6H^2 -m_1^2} \frac{m_1^2 - 6H^2}{m_1^2 -
 2H^2}  \right)\frac{1}{4}\tilde{\gamma}_{\mu\nu}T^{(+)} \right)\, .
\end{align}
To find the explicit form of the mode functions, 
we take the small $\dd/r_c^{(\pm)}$ limit keeping $m_1^2$ constant, 
which makes the other KK modes decoupled. 
In this limit we necessarily have $|\mathcal{H}|/m_1 \ll 1$ 
in the normal branch, and hence we approximate $\mathcal{H}=0$ and $a=1$.
Under this approximation, using Eqs.~\eqref{TTmass1} and \eqref{TTfcn}, 
we obtain $m_1^2=\dd^{-1}\left(1/r_c^{(+)}+1/r_c^{(-)}\right)$, and 
\begin{align} 
   u_0(Y) &= C_0\, , \\
   u_1(Y) &= C_1\left(1-Y-\frac{r_c^{(+)}}{r_c^{(-)}}Y\right)\,.
\end{align}
The normalization constants $C_0$ and $C_1$ 
are determined so that the modes are normalized 
with respect to the inner product~\eqref{inprou}. 
Keeping the leading order of $\dd/r_c^{(\pm)}$, the inner product 
\eqref{inprou} is evaluated only by the boundary contributions, 
and we have 
\begin{align} 
   C_0^2 &= \frac{1}{2} \left( r_c^{(+)} +r_c^{(-)} \right)^{-1}\, , \\
   C_1^2 &= \frac{1}{2r_c^{(+)}} \left( 1+ \frac{r_c^{(+)}}{r_c^{(-)}}
 \right)^{-1}\,  .
\end{align}
Finally, we obtain the metrics induced on the branes as
\begin{align}
   \bar h_{\mu\nu} (y_{+}) &= h^{(0)}_{\mu\nu} \left( 1+ \frac{r_c^{(-)} }{r_c^{(+)} }\right)^{-1} + h^{(1)}_{\mu\nu} \left( 1+ \left( \frac{r_c^{(-)}}{r_c^{(+)}} \right)^{-1} \right)^{-1}\,,  \nonumber \\
   \label{DGP2}
   \bar h_{\mu\nu} (y_{-}) &= h^{(0)}_{\mu\nu} \left( 1+ \frac{r_c^{(-)}
 }{r_c^{(+)} }\right)^{-1} - h^{(1)}_{\mu\nu} \left( 1+
 \frac{r_c^{(-)}}{r_c^{(+)}} \right)^{-1}\, .
\end{align}

\subsection{dRGT bigravity}

In this subsection we derive the linearized perturbation equations in dRGT bigravity around de Sitter background.
The action of dRGT bigravity is
\begin{align}
 S= \int d^4x\sqrt{-g}\left[ \frac{M_{pl}^2}{2}(R+V(g,\tilde{g}))+L_m
 \right] + \frac{\chi M_{pl}^2}{2}\int d^4x\sqrt{-\tilde{g}}\tilde{R}\, ,
\end{align}
where $g_{\mu\nu}$ and $\tilde{g}_{\mu\nu}$ are, respectively, the
physical and the hidden metrics.
$M_{pl}$ is the 4 dimensional Planck mass for $g_{\mu\nu}$ while
$\sqrt{\chi}M_{pl}$ is that for $\tilde{g}_{\mu\nu}$.
$L_m$ is the matter Lagrangian and the matter is assumed to couple only to the physical metric.
The interaction between $g_{\mu\nu}$ and $\tilde{g}_{\mu\nu}$ is given by
\begin{align} 
   V &= \sum^4_{n=0} c_n V_n\, ,
\end{align}
with
\begin{align}
   V_0 &= 1 \, , \ \ 
   V_1 = [Y] \, , \ \ 
   V_2 = [Y]^2 - [Y^2] \, , \ \ 
   V_3 = [Y]^3 - 3 [Y] [Y^2] + 2 [Y^3] \, , \\
   V_4 &= [Y]^4 -6 [Y]^2 [Y^2] + 8 [Y] [Y^3] +3 [Y^2]^2 -6 [Y^4] \, ,
\end{align}
where we have introduced $Y^{\mu}_{\nu} = \sqrt{ g^{\mu\alpha} \tilde{g}_{\alpha\nu} } $ and $[Y^n] = \mathrm{Tr}(Y^n)$.

The equations of motion of dRGT bigravity can be written as
\begin{align} \label{bi1}
&   G^{\mu}_{\nu} + B^{\mu}_{\nu} = M_{pl}^{-2} T^{\mu}_{\nu}\, , \\
   \label{bit}
&   \chi \tilde{G}^{\mu}_{\nu} + \tilde{B}^{\mu}_{\nu} = 0\, ,
\end{align}
where
\begin{align} 
&   B^{\mu}_{\nu} = m^2 \left[ V \delta^{\mu}_{\nu} - \left( V' Y \right)^{\mu}_{\nu} \right]\, , \\
&   \tilde{B}^{\mu}_{\nu} = m^2 \left( \frac{ det (\tilde{g}) }{ det (g) } \right)^{-1/2} \left( V' Y \right)^{\mu}_{\nu}\, ,
\end{align}
where $(V')^{\mu}_{\nu} = \partial V / \partial Y^{\nu}_{\mu}$.

In the case of de Sitter background, metrics are related to each other
as $\tilde{g}_{\mu\nu} = \omega^2 g_{\mu\nu}$.
For simplicity, we define functions $f(\omega)$ and $\bar{f}(\omega)$ as
\begin{align}
   f(\omega) :=& V|_{Y^{\mu}_{\nu} = \omega \delta^{\mu}_{\nu}} = c_0 + 4c_1 \omega +12c_2 \omega^2 + 24c_3 \omega^3 + 24c_4 \omega^4\, , \\
   \bar{f}(\omega)\delta^{\mu}_{\nu} :=&  \left[ V \delta^{\mu}_{\nu} - \left( V' Y \right)^{\mu}_{\nu} \right] _{Y^{\mu}_{\nu} = \omega \delta^{\mu}_{\nu}} = \left( f - \frac{\omega}{4} f' \right) \delta^{\mu}_{\nu} = \left(c_0 + 3 c_1 \omega + 6c_2 \omega^2 + 6c_3 \omega^3\right) \delta^{\mu}_{\nu}\, ,
\end{align}
where $f' = {d f}/{d \omega}$.
Denoting the curvature radius of the background physical metric as $H^{-1}$ and the cosmological constant coupled to the physical metric as $\Lambda$, 
the background equations imply 
$$
3H^2 - m^2 \bar{f} = \Lambda,
$$ 
and 
$$
3\chi \omega H^2 - m^2 f' /4 = 0.
$$
We consider linear perturbation $g_{\mu\nu} = \gamma_{\mu\nu} + h_{\mu\nu}$, $\tilde{g}_{\mu\nu} = \omega^2 (\gamma_{\mu\nu} + \tilde{h}_{\mu\nu})$, where $\gamma_{\mu\nu}$ is the de Sitter metric with the curvature radius $H^{-1}$.
From Eqs.~\eqref{bi1} and \eqref{bit}, we obtain
\begin{align} \label{h1}
&   \mathcal{E}^{\alpha\beta}{}_{\mu\nu} h_{\alpha\beta} + 3H^2 h_{\mu\nu} + \frac{m^2}{2} \Gamma \left( h^{ (m)}_{\mu\nu} - h^{(m)} \gamma_{\mu\nu}\right) = M_{pl}^{-2} T_{\mu\nu}\, , \\
\label{h2}
&   \mathcal{E}^{\alpha\beta}{}_{\mu\nu} \tilde{h}_{\alpha\beta} + 3H^2 \tilde{h}_{\mu\nu} - \frac{m^2}{2\chi \omega^2} \Gamma \left( h^{ (m) }_{\mu\nu} - h^{(m)} \gamma_{\mu\nu}\right) = 0\, ,
\end{align}
with $\Gamma := -\omega \bar{f}'/3 = c_1 \omega + 4 c_2 \omega^2 + 6 c_3 \omega^3$, $h_{\mu\nu}^{(m)} = h_{\mu\nu} - \tilde{h}_{\mu\nu}$ and
\begin{align} \label{E}
   \mathcal{E}^{\alpha\beta}{}_{\mu\nu} h_{\alpha\beta} := 
- \frac{1}{2}
 \left( \Box h_{\mu\nu} + \nabla_{\mu} \nabla_{\nu} h - \nabla_{\nu}
 \nabla_{\sigma} h^{\sigma}_{\mu} - \nabla_{\mu} \nabla_{\sigma}
 h^{\sigma}_{\nu} - \gamma_{\mu\nu} \Box h + \gamma_{\mu\nu}
 \nabla_{\alpha} \nabla_{\beta} h^{\alpha\beta} + 4H^2 h_{\mu\nu} - H^2
 h \gamma_{\mu\nu} \right) \, ,
\end{align}
where $\nabla_{\mu}$ is the covariant differentiation associated with
$\gamma_{\mu\nu}$ and $\Box=\gamma^{\mu\nu}\nabla_{\mu}\nabla_{\nu}$.
Combining Eqs.~\eqref{h1} and \eqref{h2}, we can decompose these two modes into one massless mode $h_{\mu\nu}^{(0)} = h_{\mu\nu} + \chi \omega^2 \tilde{h}_{\mu\nu}$ and one massive mode $h_{\mu\nu}^{(m)}$. 
Using the energy conservation law and the Bianchi identity, we obtain 
\begin{align} \label{ec}
   \nabla_{\mu} h^{\mu (m)}_{\nu} - \nabla_{\nu} h^{(m)}=0\, .
\end{align}
Using Eqs.~\eqref{h1}, \eqref{h2}, \eqref{E} and \eqref{ec}, the equation of motion for the massive mode becomes
\begin{align} \label{hm}
   \Box h_{\mu\nu}^{(m)} - \nabla_{\mu} \nabla_{\nu} h^{(m)} -H^2 \left(2 h_{\mu\nu}^{(m)} + \gamma_{\mu\nu} h^{(m)} \right) - m_{\rm eff}^2 \left( h^{(m)}_{\mu\nu} - \gamma_{\mu\nu} h^{(m)} \right) = - 2M_{pl}^{-2} T_{\mu\nu} \, ,
\end{align}
where 
\begin{align}\label{grmass}
m^2_{\rm eff} = m^2 \left(1 + (\chi \omega^2)^{-1}\right)\Gamma \, .
\end{align} 
Taking the trace of Eq.~\eqref{hm}, 
\begin{align} \label{trhm}
   h^{(m)} = -\frac{2M_{pl}^{-2}}{3\left( m^2_{\rm eff} -2H^2 \right)} T\, .
\end{align}
The traceless part of Eq.~\eqref{hm} leads
\begin{align} \label{trlshm}
   h_{\mu\nu}^{(m)} - \frac{1}{4} h^{(m)} \gamma_{\mu\nu} =
 \frac{-2M_{pl}^{-2}}{  \Box - 2H^2 - m_{\rm eff}^2 } \left( T_{\mu\nu}
 - \frac{1}{4} \gamma_{\mu\nu}T\right) +  \left( \nabla_{\mu}
 \nabla_{\nu} -\frac{1}{4}\gamma_{\mu\nu}\Box \right) \frac{1
}{ \Box + 6H^2 - m_{\rm eff}^2 } h^{(m)} \, ,
\end{align}
Here we used the identity for an arbitrary scalar $Y$,
\begin{align}
   \frac{1}{ \Box - 2H^2 - m_{\rm eff}^2 } \left( \nabla_{\mu} \nabla_{\nu} -\frac{1}{4}\gamma_{\mu\nu}\Box \right) Y = \left( \nabla_{\mu} \nabla_{\nu} -\frac{1}{4}\gamma_{\mu\nu}\Box \right) \frac{1}{ \Box + 6H^2 - m_{\rm eff}^2 }Y\, .
\end{align}
The term proportional to $\nabla_{\mu} \nabla_{\nu}$ in Eq.~\eqref{trlshm} can be erased by a gauge transformation.
Combining Eq.~\eqref{trlshm} and Eq.~\eqref{trhm}, we obtain
\begin{align} \label{mdRGT}
   (\Box -2H^2 - m_{\rm eff}^2) h_{\mu\nu}^{(m)} = -2M_{pl}^{-2} \left(
 T_{\mu\nu} - \left( 1 + \frac{1}{3} \frac{\Box -2H^2 -m_{\rm eff}^2}
 {\Box +6H^2 -m_{\rm eff}^2} \frac{m_{\rm eff}^2 - 6H^2}{m_{\rm eff}^2 -
 2H^2}  \right)\frac{1}{4} \gamma_{\mu\nu}T \right)\, .
\end{align}
We also find the equation for the massless mode $h_{\mu\nu}^{(0)}$ as 
\begin{align} \label{mldRGT}
    (\Box -2H^2) h_{\mu\nu}^{(0)} = -2M_{pl}^{-2} \left[ T_{\mu\nu} - \frac{1}{4}\left( 1 + \frac{\Box - 2H^2}{\Box + 6H^2} \right) \gamma_{\mu\nu}T  \right] .
\end{align}
The equations for massive and massless modes in dRGT
bigravity \eqref{mdRGT} and \eqref{mldRGT} take the same form as the equations in DGP model~\eqref{graviton} and \eqref{mgraviton}.
We can write two metrics using massless and massive gravitons as
\begin{align} 
   h_{\mu\nu} & =\frac{1}{1+\chi \omega^2} \left( h_{\mu\nu}^{(0)} + \chi \omega^2 h_{\mu\nu}^{(m)} \right) , \nonumber \\
   \label{bi2}
   \tilde{h}_{\mu\nu} & = \frac{1}{1+\chi \omega^2} \left( h_{\mu\nu}^{(0)} - h_{\mu\nu}^{(m)} \right) .
\end{align} 

To see the correspondence between the effective bigravity derived from
DGP two-brane model and dRGT bigravity, we compare Eq.~\eqref{DGP2} and
Eq.~\eqref{bi2}.
We find that the the metrics on the $(+)$-brane and the $(-)$-brane in
DGP two-brane model agree with the physical and the hidden metrics in
dRGT bigravity if we identify as $M_{pl}^{-2} = \kappa_{4(+)}^2$ and
$\chi\omega^2 = r_c^{(-)}/r_c^{(+)}$.
Notice that dRGT bigravity accepts the scale transformation $\tilde{g}_{\mu\nu}
\rightarrow \Omega^2 \tilde{g}_{\mu\nu}$, $\chi \rightarrow
\chi/\Omega^2$, $c_i \rightarrow c_i/\Omega^i$.
The metric perturbations $h_{\mu\nu}$ and $\tilde{h}_{\mu\nu}$ are
invariant under this transformation, which enables us to arbitrarily 
rescale the hidden metric $\tilde{g}_{\mu\nu}$.

Let us consider a homogeneous and isotropic perturbation in DGP two-brane model 
and compare it with the result in ref.~\cite{GW}.
We can easily find that the light cone of the metric on $(-)$-brane changes 
in the same way as that of the hidden metric $\tilde{c}-1$ in ref.~\cite{GW}, 
because the induced metric perturbation is identical. 
In order to see the nonlinear effect by raising the background energy scale $H$ as 
anticipated, we would be able to compare the Friedmann equations.
However, it is hard to obtain the Friedmann equation in DGP two-brane model 
because of the back reaction on the structure of the fifth dimension 
caused by changing the brane cosmological constant. 
Furthermore, we find that there is a tachyonic instability 
in the sector of the stabilization scalar field, 
which will be shown in the succeeding section, 
and hence we will not further discuss the non-linear perturbation in this paper.

\section{Instability}

In this section we investigate instabilities in DGP two-brane model with
a scalar field for the radion stabilization and those in dRGT bigravity.
We will discuss how one can interpret the instabilities in dRGT bigravity 
from the viewpoint of higher dimensional gravity and the difference in 
the way how instabilities appear between these two models.

\subsection{DGP model}

It is known that the self-accelerating branch of DGP two-brane model 
inevitably has ghost~\cite{Izumi}.
Therefore, here we consider the normal branch, which is obtained by 
a continuous deformation of the model 
from a rather simple setup that accepts 
the solution with two Minkowski branes and the bulk satisfying $1\mp2
r_c^{(\pm)} \mathcal{H}_{\pm}>0$. 

Using the eigenfunctions $u_i$ and $v_i$, we can obtain the solution for
the perturbation $h_{\mu\nu}^{(TT)}$ and $\phi$ induced by the matter on
the $(+)$-brane. 
As was discussed in \cite{Izumi}, a KK graviton 
in the mass range $0<m^2<2H^2$ and a spin-0 mode with the mass below $-4H^2$ 
become a Higuchi ghost and a scalar ghost, respectively, 
 and lead to quantum instabilities. 
We begin with Minkowski brane solution and consider to gradually
increase the brane tension of the $(+)$-brane. 
As we will see below, one can verify that on the background 
solution obtained in this manner ghost does not arise
as long as $1\mp2 r_c^{(\pm)}\mathcal{H} >0$, 
thanks to the identity~\eqref{id}.
When only small perturbation from this solution is concerned, 
the mass spectrum will remain to be essentially the same,  
and therefore all KK graviton masses are above $2H^2$ and 
all spin-0 mode masses are above $0$. Hence, 
there appears no instability.
If we continue to increase the brane tension further, 
we can obtain de Sitter branes with the four-dimensional 
Hubble parameter $H$.
The mass eigenvalues vary but they must satisfy Eq.~\eqref{id}.
The assumption $\epsilon^{(\pm)} \geq 0$ guarantees that the 
norm of $v_i$ is always positive, and hence $v_i^2$ is positive. 
By assumption, the solution remains in the normal branch, 
in which $1-2r_c^{(+)} \mathcal{H} _+>0$ is satisfied. 
In this case, when $m_i^2$ 
crosses the critical value $2H^2$ from above, 
the first term on the l.h.s.~in Eq.~\eqref{id} 
diverges to positive infinity.   
Similarly, when  
$\mu_i^2$ crosses $-4H^2$, 
the first term in the square brackets on the l.h.s.~diverges 
to positive infinity.   
Whilst, there is no other term that 
diverges to negative infinity.
Therefore the crossing of the critical masses 
violates the identity~\eqref{id}, and hence it never happens. 
Hence, the mass spectrum on the background solution 
constructed in this manner is guaranteed to be free from ghost. 

However, Eq.~\eqref{bgjc} implies that, as we increase the brane
tension, $|\mathcal{H}_{\pm}|$ becomes as large as $r_c^{(\pm)}H^2$
unless miraculous cancellation by the change of $V_{(\pm)}(\psi_{\pm})$
occurs.  
At least, as long as we consider models in which $\psi_{\pm}$ is pinned
down by sufficiently large $V''_{(\pm)}$, this cancellation cannot be
expected.
Therefore, when $r_c^{(\pm)}$ is large, $|\mathcal{H}_{\pm}|$ becomes
larger than $\mathcal{O}(1/r_c^{(\pm)} )$ and the lowest scalar mass
squared $\mu^2$ crosses $-4H^2$ just by considering a slightly higher
energy regime, $H \gtrsim 1 / r_c^{(\pm)}$.
As we increase the brane tension little by little starting with
Minkowski branes, $\mu^2$ becomes negative and the scalar tachyon
appears before $\mu^2$ crosses $-4H^2$. 
This tachyonic instability would mean the boundary whether the spacetime
structure with stably separated two branes is sustainable or not.
In the presence of the scalar tachyon, two branes will move away from
each other further and further, and hence DGP model cannot reproduce the
bigravity.
To avoid such an instability in the high energy regime, we should invent
some other mechanism to stabilize the separation between two branes much
more strongly, which is left for future work.
Once either of $1\mp2 r_c^{(\pm)} \mathcal{H}_{\pm}$ becomes
negative, the solution is in the self-accelerating branch.
Izumi et al.~\cite{Izumi} proved that we cannot avoid the ghost
instability in this branch using the same identity \eqref{id}, which we 
used to prove the absence of instability in the normal branch. 
If we further deform the model, we may have a transition between a
scalar mode with $\mu^2 < - 4H^2$ and Higuchi ghost, which was discussed
in Ref.~\cite{Izumi}.

\subsection{dRGT bigravity model}

Here we study how ghost appears in dRGT bigravity model.
We consider FLRW background and its perturbation in dRGT bigravity.
We assume the background geometry as given by 
\begin{align} 
 &  ds^2 = a(t)^2 (-dt^2 + d\vec{x}^2)\, , \\
 &  \tilde{ds}^2 = \tilde{a}(t)^2 ( - \tilde{c}^2 (t) dt^2 + d\vec{x}^2)\, .
\end{align}
Following the discussion in \cite{GW}, we select the healthy branch
$\tilde{c}aH-\tilde{a}' / \tilde{a} = 0$ to solve the conservation
equation $\nabla^{\mu}B_{\mu\nu} = 0$. 
the equation of motion becomes
\begin{align} 
   3H^2 = \kappa_4^2 (\rho_m + \rho_V)\, ,  
\end{align}
where $H:=\partial_t a/a^2$, $\rho_m$ is the matter energy density and
\begin{align}
   \frac{\kappa_4^2}{m^2} \rho_V &= \bar{f}(\omega)\, , \\
   \label{matter}
   \frac{\kappa_4^2}{m^2} \rho_m &= \frac{c_1}{\chi \omega} + \left( \frac{6c_2}{\chi} - c_0 \right) + \left( \frac{18c_3}{\chi} - 3c_1 \right)\omega + \left( \frac{24c_4}{\chi} - 6c_2 \right) \omega^2 - 6c_3\omega^3 \nonumber \\
   &= -\bar{f}(\omega) + \frac{1}{4\chi \omega} f'(\omega) =: F(\omega)\, ,
\end{align}
where $\omega := \tilde{a}/a$. 
The graviton mass squared $m_{\rm eff}^2$ is given in Eq.~\eqref{grmass}
and positive when $\Gamma(\omega)>0$.
According to \cite{cos5}, the helicity-0 mode of graviton becomes ghost when $m_{\rm eff}^2$ is smaller than $2H^2$, which is the so-called Higuchi ghost. 
After some calculation, we find
\begin{align}  
   m_{\rm eff}^2 - 2H^2 = -\frac{m^2 \omega}{3} F'(\omega)\, .
\label{Higuchibound}
\end{align}
Therefore, we can judge the appearance of ghost by the sign of $F'(\omega)$.
When $F'(\omega)$ is negative, the solution is free from the Higuchi ghost.

Suppose that the vacuum energy is tuned so as to possess a vacuum
Minkowski solution with $\omega = \omega_0$ and $\rho_m(\omega_0) = 0$.  
We consider a branch of the solution that evolves to this Minkowski solution.
As we increase the energy density $\rho_m$, FLRW solution ceases to exist at the point where $F'(\omega) = 0$.
Up to this energy density, $F'(\omega)$ remains to be negative as far as $m_{\rm eff}^2>0$ is satisfied for $\omega = \omega_0$.
Equation~\eqref{Higuchibound} tells that 
$m_{\rm eff}^2 - 2H^2$ is kept to be positive on this branch.
Therefore, we can conclude that the cosmological solution constructed in this way has no Higuchi ghost as long as the positivity of the graviton mass squared $m^2_{\rm eff}$ in the Minkowski limit is guaranteed.
On the branch with $F'(\omega)>0$, the graviton's mass $m^2_{\rm eff}$ is less than $2H^2$ and the Higuchi ghost appears.

The way how ghost appears in dRGT model is clearly different from the
case in DGP model. 
In the dRGT bigravity, before the Higuchi ghost appears, 
there is no instability, while in DGP model the onset of the 
instability is tachyonic. 
We can understand this disagreement as caused by the truncation of the
scalar mode.
On one hand, in DGP two-brane model the scalar modes could be effectively
neglected when all of them are sufficiently heavy.
However, once some of them become light, we cannot neglect them any
more. Hence, we found a tachyonic instability. 
On the other hand, in dRGT bigravity scalar modes do not exist from the
beginning.
Therefore, the appearance of the tachyonic scalar mode observed in DGP
two-brane model does not show up in the corresponding dRGT bigravity,
which keeps dRGT bigravity ``healthy'' even in the energy regime higher than
$1/r_c^{(\pm)}$.

\section{Summary}

In this paper, we have investigated whether or not dRGT bigravity
can be embedded in higher dimensional gravity. 
Here, we have considered DGP two-brane model 
with sufficiently large brane induced gravity terms and  
Goldberger-Wise radion stabilization. 
We have chosen the model parameters so as to accept 
background solutions in which the brane separation 
is sufficiently small compared with the length scale 
determined by the ratio between 
five-dimensional and four-dimensional Newton's constants. 
We have proved that DGP two-brane model in such regime 
can reproduce bigravity as its 
low energy effective theory with the help of the stabilization
mechanism. Namely, almost massless degrees of freedom in 
the gravity sector are composed of one massless graviton and 
one massive graviton with a small mass. 

DGP model is known to have the normal branch and the self-acceleration
branch. We clearly identify the condition to distinguish these two 
branches in the setup with a scalar field introduced for the 
purpose of radion stabilization. 
We succeeded in proving that 
the model does not have ghost as long as the normal branch solution is
chosen. 
Putting aside the issue of Higuchi ghost, 
since DGP two-brane model does not have ghost in the scalar-type 
perturbation irrespective of the choice of branch, the ghost 
corresponding to BD ghost in bigravity is guaranteed to be absent. 
Therefore, the low energy effective theory of DGP two-brane model is 
expected to be identical to dRGT bigravity, which is the unique bigravity 
theory that is free from BD ghost. As is expected, we also succeeded 
in proving this identity, at least, at the linear level.

We have also studied how ghost appears in DGP two-brane model and dRGT
bigravity when we continuously modify the model parameters. 
In both models we can consider backgrounds that are free from ghost 
at low energies. 
In DGP two-brane model with stabilization,
however, it is difficult to avoid ghost when we slightly increase 
the background energy scale. 
This is because the stabilization of the 
brane separation is hard to maintain as long as we keep 
the conditions for the normal branch. As a result, 
a tachyonic four-dimensional scalar mode arises. 
By contrast, in dRGT bigravity such a four-dimensional scalar degree of 
freedom corresponding to the brane separation 
does not exist from the beginning, and hence 
the model remains free from the instability. 
Therefore, it turned out that the correspondence between DGP
two-brane model with scalar-field stabilization mechanism and dRGT
bigravity holds only in the very limited low energy regime.

Unfortunately, because of this instability that occurs 
at a relatively low energy, 
it is difficult to fully justify investigating the 
properties of dRGT bigravity by using the counterpart in the braneworld 
setup. Nevertheless, it is suggestive to point out that 
the Vainshtein mechanism 
in the low energy regime of dRGT bigravity 
explored in Ref.~\cite{GW} tells that 
the physical and hidden metrics are similarly excited near 
gravity sources. It might be natural to expect that 
the same feature will arise for the metrics induced on 
both branes when the brane separation is very small. 
The effective gravitational coupling that appears in the effective 
Friedmann equation and the local Newton's law within the 
Vainshtein radius in the low energy regime of dRGT bigravity 
is given by the sum of the four-dimensional Planck mass squared 
for the physical and the hidden metrics. This feature can also 
be understood as the dilution of gravitational force line,  
which is very familiar in the braneworld context. 
In our future publication we will study 
whether or not there is more efficient stabilization 
mechanism that maintains the correspondence
even in the higher energy regime.

Acknowledgements
This work was supported in part by the Grant-in-Aid for Scientific
Research (Nos. 21244033, 
21111006, 24103006 and 24103001). We also would like to mention 
that the discussion during 
the molecule-type YITP workshop: YITP-T-13-08 was useful to complete this
work.

\end{document}